\documentclass[12pt]{revtex4}
\usepackage{amssymb,amsmath,graphicx,latexsym}
\usepackage{bm}
\usepackage{epsfig}

\begin{document}

\title{Power-law and intermediate inflationary models in $f(T)$-gravity}

\author{K. Rezazadeh$^{1}$\footnote{rezazadeh86@gmail.com}, A. Abdolmaleki$^{2}$\footnote{aabdolmaleki@uok.ac.ir} and K. Karami$^{1}$\footnote{kkarami@uok.ac.ir}}
\address{$^{1}$Department of Physics, University of Kurdistan, Pasdaran St., Sanandaj, Iran\\
$^2$Research Institute for Astronomy $\&$
Astrophysics of Maragha (RIAAM), P.O. Box 55134-441, Maragha, Iran}

\begin{abstract}
We study inflation in the framework of $f(T)$-gravity in the presence of a canonical scalar field. After reviewing the basic equations governing the background cosmology in $f(T)$-gravity, we turn to study the cosmological perturbations and obtain the evolutionary equations for the scalar and tensor perturbations. Solving those equations, we find the power spectra for the scalar and tensor perturbations. Then, we consider a power-law $f(T)$ function and investigate the inflationary models with the power-law and intermediate scale factors. We see that in contrast with the standard inflationary scenario based on the Einstein gravity, the power-law and intermediate inflationary models in $f(T)$-gravity can be compatible with the observational results of Planck 2015 at 68\% CL. We find that in our $f(T)$ setting, the potentials responsible for the both power-law and intermediate inflationary models have the power-law form $V(\phi ) \propto {\phi ^m}$ but the power $m$ is different for them. Therefore, we can refine some of power-law inflationary potentials in the framework of $f(T)$-gravity while they are disfavored by the observational data in the standard inflationary scenario. Interestingly enough, is that the self-interacting quartic potential $V(\phi ) \propto {\phi ^4}$ which has special reheating properties, can be consistent with the Planck 2015 data in our $f(T)$ scenario while it is ruled out in the standard inflationary scenario.
\\
\end{abstract}

\maketitle


\section{Introduction}\label{secintr}

Inflationary scenario was proposed to overcome some of the basic problems of the Hot Big Bang cosmology such as the flatness problem, the horizon problem and also the magnetic monopole problem \cite{Sta80, Gut81, Lin82, Alb82, Lin83, Lin86a, Lin86b}. In addition, growth of the perturbations seeded during inflationary era can successfully explain the large-scale structure (LSS) formation as well as the anisotropy observed in the cosmic microwave background (CMB) radiation \cite{Muk81, Haw82, Sta82, Gut82}. Therefore, applying the experimental results from LSS and CMB radiation, we are able to obtain useful information about the inflationary stage of the universe. Important observational results are represented by the Planck 2015 collaboration \cite{Pla15} that they are obtained from probing of the CMB radiation anisotropies in both temperature and polarization. Using these observational results, we can distinguish viable inflationary models and also constrain them.

In inflation theory, a rapid accelerating expansion is considered before the radiation dominated era. In the standard inflationary scenario, a canonical scalar field is regarded in the framework of Einstein's general relativity (GR) to explain the accelerating expansion of the inflationary era. Viability of different inflationary models in the framework of standard inflationary scenario in light of observational results has been extensively investigated in the literature \cite{Mar14, Rez15a, Oka14, Hos14a, Hos14b, Hos14c, Gen15}. However, there are other inflationary models represented on the base of extended theories of gravity. One important class of this category includes the models based on $f(R)$-gravity in which the Ricci curvature scalar $R$ in the action is replaced by an arbitrary function of $f(R)$ \cite{Buc70}. The well-known instance for this class is the Starobinsky $R^2$ inflation \cite{Sta80} which is the first inflationary model and it is based on addition of the term $R^2$ to the Einstein-Hilbert term $R$ in the action. Although this model is the first inflationary model, it is in well agreement with the experimental data as it has been demonstrated by the Planck 2015 collaboration \cite{Pla15}. In order to find other inflationary models in the framework of $f(R)$-gravity see \cite{Pi09, Hua14, Rin15, Art15, Seb15}.

Another important class of the inflationary models based on the extended theories of gravity includes the models founded on the teleparallel gravity (TG) and its extension, $f(T)$-gravity. TG was originally proposed by Einstein \cite{Ein30} in an attempt of unifying gravity and electromagnetism. Later, Einstein left TG because it failed in the attempt of this unification and also the curvature tensor of the Weitzenbock connection vanishes. It has been shown that TG can provide an alternative for GR \cite{Hay79}. Then, the idea of the teleparallel equivalent to general relativity (TEGR) was developed. Subsequently, the TEGR was generalized to $f(T)$-gravity by replacing a general $f(T)$ function instead of the torsion scalar $T$ in the action \cite{Fer07, Fer08}. The basic variables in $f(T)$-gravity are the tetrad fields $e_{i\mu}$ where the Weitzenbock connection instead of the Levi-Civita connection is used to define the covariant derivative. Consequently, the spacetime has no curvature but contains torsion. The main advantage of $f(T)$ theory is the fact that its field equations are second order which are significantly simpler than the fourth order equations of $f(R)$-gravity \cite{Wu10a, Wu11, Wei11, Kar13a, Kar13b}. Although, the models based on $f(T)$-gravity can be regarded as an alternative to $f(R)$ theories \cite{Ben09}, in contrast with $f(R)$ scenario, $f(T)$-gravity is not dynamically equivalent to teleparallel action plus a scalar field via conformal transformation \cite{Yan11a}.

Recently, $f(T)$-gravity has aroused a great interest in cosmological applications. At first, $f(T)$-gravity was proposed as models for inflation \cite{Fer07, Fer08}. Then, models based on $f(T)$-gravity was considered to describe the present accelerating expansion of the universe without resorting to dark energy (DE) \cite{Wu10a, Wu11, Wei11, Kar13a, Kar13b, Ben09, Yan11a, Yan11b, Wu10b, Lin10, Ben11, Myr11, Bam11a, Zha11}. Also, thermodynamics of $f(T)$-gravity models has been investigated in \cite{Bam11b, Mia11, Kar12}. Reconstructing of $f(T)$ theories equivalent to models based on scalar fields is subject of \cite{Dao12}. LSS formation in the framework of $f(T)$-gravity has been regarded in \cite{Li11, Nes13}. Cosmological perturbations in $f(T)$-gravity has been studied in \cite{Den11, Che11, Zhe11, Cai11, Izu13, Beh15a, Beh15b}. Also, recently, some inflationary models in the framework of $f(T)$-gravity have been investigated in \cite{Nas14a, Nas14b, El15, Jam15}.

In the present work, we focus on the study of inflation in the framework of $f(T)$-gravity in the presence of a canonical scalar field. We choose a power-law form for $f(T)$ function in the action and then investigate inflationary models with the power-law and intermediate scale factors in our setting. The power-law inflation is specified by the scale factor $a(t) \propto {t^q}$ where $q>1$, and in the standard inflationary scenario based on a canonical scalar field in the framework of Einstein gravity, it is driven by an exponential potential which is not consistent with the observational results, as it has been shown in \cite{Rez15a, Unn13}. Besides, the intermediate inflation is characterized by the scale factor $a(t) \propto \exp \left[ {A{{\left( {{M_P}t} \right)}^\lambda }} \right]$ where $A>0$ and $0<\lambda<1$ and it arises from an inverse power-law potential in the standard inflationary scenario \cite{Bar06, Bar07, Rez15b}, that it is not favored by the Planck 2015 data, as it has been discussed in \cite{Rez15a, Rez15b}. In the present work, our main goal is to refine these inflationary models in light of Planck 2015 results in the framework of $f(T)$-gravity. To this aim, first we review the background cosmological consequences of $f(T)$-gravity and we will consider them in the slow-roll approximation to find the inflationary potentials responsible for these models. Then, we will study the cosmological perturbations to obtain the power spectra for the scalar and tensor perturbations. This makes it possible for us to check the viability of our model in comparison with the observational data.

This paper is organized as follows. In section \ref{secfT}, we review the basics equations for the cosmological background evolution in $f(T)$-gravity. In section \ref{secper}, we study the cosmological perturbations theory in $f(T)$-gravity and obtain the scalar and tensor power spectra. Then, in sections \ref{secpl} and \ref{secint}, we investigate the power-law and intermediate inflationary models, respectively, in the framework of $f(T)$-gravity. Finally in section \ref{seccon}, we summarize the concluding remarks.

\section{$f(T)$-gravity}\label{secfT}

In this section, we provide a brief review on the basic equations for background cosmological evolution in $f(T)$-gravity. Generalizing the TEGR, we consider the action of $f(T)$-gravity as \cite{Fer07, Fer08}
\begin{equation}\label{action}
I = \frac{{M_P^2}}{2}\int {{d^4}x\,e\,\left[ {f(T) + {L_\phi}} \right]},
\end{equation}
where ${M_P} = 1/\sqrt {8\pi G}$ is the reduced Planck mass and $L_{\phi}$ is the scalar field Lagrangian. Also, $e = \det (e_\mu ^i) = \sqrt { - g}$ where $e_\mu ^i$ is the vierbein field which is used as a dynamical object in TG.

For a spatially flat FRW universe, the modified Friedmann equations in $f(T)$-gravity read \cite{Wu10a, Wu11, Wei11, Kar13a, Kar13b}
\begin{eqnarray}
\label{Fri1}
{H^2} &=& \frac{1}{{3M_P^2}}\left( {{\rho _T} + {\rho _\phi }} \right),
\\
\label{Fri2}
\dot H + \frac{3}{2}{H^2} &=&  - \frac{1}{{2M_P^2}}\left( {{p_T} + {p_\phi}} \right),
\end{eqnarray}
where $H \equiv \dot a/a$ is the Hubble parameter and $\rho_\phi$ and $p_\phi$ stand for the energy density and pressure of the scalar field, respectively. Also, $\rho_T$ and $p_T$ are the energy density and pressure due to the torsion contribution, respectively, and they are defined as
\begin{eqnarray}
\label{rhoT}
{\rho _T} &=& \frac{{M_P^2}}{2}\left( {2T{f_{,T}} - f - T} \right),
\\
\label{pT}
{p_T} &=&  - \frac{{M_P^2}}{2}\left[ { - 8\dot HT{f_{,TT}} + \left( {2T - 4\dot H} \right){f_{,T}} - f + 4\dot H - T} \right].
\end{eqnarray}
Here $f_{,T}=df/dT$. For a spatially flat FRW metric, the torsion scalar has a relation with the Hubble parameter as \cite{Wu10a, Wu11, Wei11, Kar13a, Kar13b}
\begin{equation}\label{T}
T =  - 6{H^2}.
\end{equation}
Note that in the case of $f(T)=T$, Eqs. (\ref{rhoT}) and (\ref{pT}) yield $\rho_T=0$ and $p_T=0$ so that Eqs. (\ref{Fri1}) and (\ref{Fri2}) transform to the usual Friedmann equations in the TEGR.

The torsion and scalar field energy densities, independently, satisfy the conservation equations as
\begin{eqnarray}
\label{rhodotT}
{\dot \rho _T} + 3H\left( {{\rho _T} + {p_T}} \right) &=& 0,
\\
\label{rhodotphi}
{\dot \rho _\phi} + 3H\left( {{\rho _\phi} + {p_\phi}} \right) &=& 0.
\end{eqnarray}

In this paper, we assume that the matter content of the universe to be a canonical scalar field. Therefore, the energy density and pressure of the scalar filed, respectively, are given by
\begin{eqnarray}
\label{rhophi}
{\rho _\phi } &=& \frac{1}{2}{\dot \phi ^2} + V(\phi ),
\\
\label{pphi}
{p _\phi } &=& \frac{1}{2}{\dot \phi ^2} - V(\phi ),
\end{eqnarray}
where ${\dot \phi ^2}/2$ and $V(\phi)$ are the kinetic energy and potential of the scalar field, respectively. Substituting $\rho_\phi$ and $p_\phi$ from the above equations into the conservation equation (\ref{rhodotphi}) leas to the evolution equation of the scalar field as
\begin{equation}\label{phiddot}
\ddot \phi  + 3H\dot \phi  + {V_{,\phi }} = 0,
\end{equation}
where $V_{,\phi}=dV/d\phi$. Notice that in $f(T)$-gravity, the set of equations containing the Friedmann equations (\ref{Fri1}) and (\ref{Fri2}), and the evolution equation governing the scalar field (\ref{phiddot}) are not independent of each other. Taking the time derivative of Eq. (\ref{Fri1}) and using (\ref{phiddot}), one can get the second Friedmann equation (\ref{Fri2}). Also, from definitions (\ref{rhoT}) and (\ref{pT}) one can obtain Eq. (\ref{rhodotT}). In what follows, we take the set of Eqs. (\ref{Fri1}) and (\ref{phiddot}), which can uniquely determine the dynamics of the universe.

In the present work, we are interested in investigating inflation in the framework of $f(T)$-gravity. To this aim, it is useful to define the Hubble slow-roll parameters as
\begin{eqnarray}
\label{eps1}
\varepsilon_1  &\equiv&  - \frac{{\dot H}}{{{H^2}}},
\\
\label{epsip1}
\varepsilon_{i+1}  &\equiv&  \frac{{\dot \varepsilon_i }}{{H\varepsilon_i}}.
\end{eqnarray}
From definition (\ref{eps1}) it is evident that in order to have inflation ($\ddot a > 0$), we should have $\varepsilon_1 < 1$. Therefore, dependent on whether the first Hubble slow-roll parameter $\varepsilon_1$ be a decreasing function or an increasing function during inflation, we can use the relation $\varepsilon_1 = 1$ to determine the initial time or the end time of inflation, respectively \cite{Zha14}.

The scalar field responsible for inflation is called ``inflaton''. During the inflationary era, variation of the inflaton $\phi$ is very slow. Also, during this era, the Hubble parameter $H$ changes slowly so that we have a quasi-de Sitter expansion. These facts allow us to apply the slow-roll conditions ${\dot \phi ^2} \ll V(\phi )$ and $\big| {\ddot \phi } \big| \ll \big| {3H\dot \phi } \big|,\,\big| {{V_{,\phi }}} \big|$ in study of inflation. In the slow-roll approximation, Eqs. (\ref{Fri1}), (\ref{rhoT}) and (\ref{phiddot}) can be combined to give
\begin{eqnarray}
\label{V}
V &=& \frac{{M_P^2}}{2}\left( {f - 2T{f_{,T}}} \right),
\\
\label{phidot}
{\dot \phi ^2} &=&  - 2M_P^2\dot H\left( {{f_{,T}} + 2T{f_{,TT}}} \right).
\end{eqnarray}
Note that for a given $f(T)$ and scale factor $a(t)$, with the help of Eqs. (\ref{V}) and (\ref{phidot}), one can find the evolutionary behaviors of the inflationary potential $V$ and the inflaton $\phi$ with respect to the cosmic time $t$. Then, one can combine the results to specify the inflationary potential $V(\phi)$.

\section{Cosmological perturbations in $f(T)$-gravity}\label{secper}

In this section, we focus on the study of cosmological perturbations in the framework of $f(T)$-gravity when a canonical scalar field is present. We work in the longitudinal gauge which only involves scalar-type metric fluctuations as \cite{Muk93}
\begin{equation}\label{ds2}
d{s^2} = \left( {1 + 2\Phi } \right)d{t^2} - {a^2}(t)\left( {1 - 2\Psi } \right)d{x^2}.
\end{equation}
Here, as usual, two functions $\Phi$ and $\Psi$ are used to characterize the scalar perturbations of the metric. We assume that the anisotropic stress vanishes and thus $\Psi  = \Phi $ which is widely found in the standard theory of cosmological perturbations \cite{Muk93}. If we combine the perturbation equations obtained in \cite{Cai11}, then we can get the complete form of the equation of motion for one Fourier mode $\Phi_k$ with the comoving wavenumber $k$ as
\begin{equation}\label{Phikddot1}
{\ddot \Phi _k} + \alpha {\dot \Phi _k} + {\mu ^2}{\Phi _k} + c_s^2\frac{{{k^2}}}{{{a^2}}}{\Phi _k} = 0,
\end{equation}
where the functions $\alpha$, $\mu$ and $c_s$ are respectively the frictional term, the effective mass, and the sound speed parameter and they are defined as \cite{Cai11}
\begin{eqnarray}
\label{alpha}
\alpha  &=& 7H + \frac{{2{V_{,\phi }}}}{{\dot \phi }} - \frac{{36H\dot H\left( {{f_{,TT}} - 4{H^2}{f_{,TTT}}} \right)}}{{{f_{,T}} - 12{H^2}{f_{,TT}}}},
\\
\label{mu}
{\mu ^2} &=& 6{H^2} + 2\dot H + \frac{{2H{V_{,\phi }}}}{{\dot \phi }} - \frac{{36H\dot H\left( {{f_{,TT}} - 4{H^2}{f_{,TTT}}} \right)}}{{{f_{,T}} - 12{H^2}{f_{,TT}}}},
\\
\label{cs}
c_s^2 &=& \frac{{{f_{,T}}}}{{{f_{,T}} - 12{H^2}{f_{,TT}}}}.
\end{eqnarray}
Moreover, if we use the Friedmann equation (\ref{Fri2}) and the evolution equation (\ref{phiddot}) for the scalar field, then we can rewrite Eq. (\ref{Phikddot1}) as
\begin{equation}\label{Phikddot2}
{\ddot \Phi _k} + \left( {H - \frac{{\ddot H}}{{\dot H}}} \right){\dot \Phi _k} + \left( {2\dot H - \frac{{H\ddot H}}{{\dot H}}} \right){\Phi _k} + \frac{{c_s^2{k^2}}}{{{a^2}}}{\Phi _k} = 0.
\end{equation}
This is the equation of motion for the gravitational potential $\Phi$ in $f(T)$-gravity in the presence of a canonical scalar field. We see that this equation is identical with the one in the standard Einstein gravity \cite{Muk93}, except the new sound speed parameter $c_s$ has been introduced.

In order to examine the evolution of perturbations, it is appropriate to work with gauge-invariant variables in order to the result be independent of the coordinate system. In the theory of cosmological perturbations, we often use the gauge-invariant variable $\zeta$ denoting the curvature perturbation in comoving coordinates, to specify the cosmological inhomogeneities \cite{Muk93}. Following \cite{Cai11}, we assume that the form of $\zeta$ is the same as that defined in the standard cosmological perturbation theory that it is given by
\begin{equation}\label{zeta}
\zeta  = \Phi  - \frac{H}{{\dot H}}\big( {\dot \Phi  + H\Phi } \big).
\end{equation}
Using the above equation together with Eq. (\ref{Phikddot2}), we can obtain
\begin{equation}\label{zetakdot}
{\dot \zeta _k} = \frac{H}{{\dot H}}\frac{{c_s^2{k^2}}}{{{a^2}}}{\Phi _k}.
\end{equation}
For the case of a generic expanding universe ${\dot \zeta _k}$ approaches zero at large length scales, $k \to 0$, because the dominant mode of ${\dot \Phi _k}$ is approximately constant. Now, we define the canonically normalized variable
\begin{equation}\label{v}
v = z_s \zeta,
\end{equation}
where
\begin{equation}\label{zs}
z_s = a\sqrt {2\varepsilon_1} M_P,
\end{equation}
and $\varepsilon_1$ is the first Hubble slow-roll parameter defined in Eq. (\ref{eps1}). Using Eqs. (\ref{zeta}), (\ref{zetakdot}), (\ref{v}) and (\ref{zs}), we reach the equation of motion for the scalar perturbations as
\begin{equation}\label{d2vk}
v_k'' + \left( {c_s^2{k^2} - \frac{{z_s''}}{z_s}} \right){v_k} = 0,
\end{equation}
where the prime denotes the derivative with respect to the conformal time $\tau  \equiv \int {dt/a} $. If the sound speed is equal to the light speed, i.e. $c_s=1$, then the above equation becomes the well-known ``Mukhanov-Sasaki equation'' governing the evolution of scalar perturbations in the standard Einstein gravity \cite{Muk93}. The above equation is similar to the equation of motion for the scalar perturbations in the $k$-inflation scenario in which a non-canonical kinetic term in the action drives an inflationary evolution \cite{Arm99}. This similarity provides us to follow the approach applied in \cite{Gar99} to solve Eq. (\ref{d2vk}) and find the spectrum of the variable $\zeta  = v/z_s$. This variable can be used to describe the scalar perturbations since it is directly related to the gravitational potential $\Phi$ by Eq. (\ref{zeta}). The gravitational potential $\Phi$ in turn is related to the scalar perturbations that lead to the LSS formation and the fluctuations of the CMB radiation temperature. Specifically, the fluctuations of the CMB radiation temperature in large angular scales are expressed as $\delta \mathcal{T}/\mathcal{T} \approx \Phi /3$ \cite{Muk93}.

During slow-roll inflation, the Hubble rate $H$, the sound speed $c_s$ and the first Hubble slow-roll parameter $\varepsilon_1$ change much slower than the scale factor $a$. Thus, from (\ref{zs}) we have $z_s''/{z_s} \approx a''/a \approx 2{(aH)^2}$. At sufficiently early times, the physical wavelength of the perturbation $a/k$ is much smaller than the ``sound horizon'' $c_s H^{-1}$ and hence the short wavelength limit condition ${c_s}k \gg aH$ is valid. In this limit, we can neglect the term $z_s''/{z_s}$ versus the term $c_s^2 k^2$ in Eq. (\ref{d2vk}). Also, we note that the fluctuation corresponds to a free scalar propagating in a flat spacetime, and naturally the initial condition takes the form of the Bunch-Davies vacuum \cite{Bun78}. Therefore, Eq. (\ref{d2vk}) can be easily solved to give the short wavelength solution
\begin{equation}\label{vk1}
{v_k} = \frac{{{{\mathop{\rm e}\nolimits} ^{ - i c_s k\tau }}}}{{\sqrt {2c_s k} }},\,\,\,\,\,\,\,\,\,\, ({c_s}k \gg aH).
\end{equation}
On the other hand, when the perturbations cross the sound horizon outward it, the term $z_s''/{z_s}$ begins to dominate over the term $c_s^2 k^2$ in Eq. (\ref{d2vk}). Consequently, Eq. (\ref{d2vk}) gives rise to the long wavelength solution
\begin{equation}\label{vk2}
{v_k} = {C_k}z,\,\,\,\,\,\,\,\,\,\, ({c_s}k \ll aH),
\end{equation}
where $C_k$ is a constant. To find the constant $C_k$, following \cite{Gar99}, we match the solutions (\ref{vk1}) and (\ref{vk2}) at the moment of the sound horizon exit for which $c_s k = aH$, and consequently we will have
\begin{equation}\label{Ck}
{\left| {{C_k}} \right|^2} = \frac{1}{{2{c_s}kz_s^2}}.
\end{equation}
Therefore, using Eqs. (\ref{v}) and (\ref{zs}), we obtain the power spectrum for the scalar perturbations in the framework of $f(T)$-gravity as
\begin{equation}\label{Ps}
{{\cal P}_s} \equiv \frac{{{k^3}}}{{2{\pi ^2}}}{\left| \zeta  \right|^2} = {\left. {\frac{{{k^3}}}{{2{\pi ^2}}}\frac{{{{\left| {{v_k}} \right|}^2}}}{{z_s^2}}} \right|_{ {c_s}k=aH}} = {\left. {\frac{{{H^2}}}{{8{\pi ^2}M_P^2c_s^3{\varepsilon _1}}}} \right|_{ {c_s}k=aH}},
\end{equation}
which should be evaluated at the sound horizon exit specified by $c_s k=a H$. This equation reduces to the standard result for slow-roll inflation if the sound speed is equal to the light speed ($c_s=1$).

The scalar spectral index is defined as
\begin{equation}\label{ns}
{n_s} - 1 \equiv \frac{{d\ln {{\cal P}_s}}}{{d\ln k}}.
\end{equation}
Since during slow-roll inflation, the Hubble parameter $H$ and the sound speed $c_s$ are almost constant, therefore using the relation $c_s k=a H$ that is valid at the sound horizon exit, we can obtain the relation
\begin{equation}\label{dlnk}
d\ln k \approx Hdt.
\end{equation}
Using Eqs. (\ref{eps1}), (\ref{epsip1}), (\ref{Ps}), (\ref{ns}) and (\ref{dlnk}), we can obtain
\begin{equation}\label{ns2}
{n_s} = 1 - 2{\varepsilon _1} - {\varepsilon _2} - 3{\varepsilon _{s1}},
\end{equation}
where we have defined the sound speed slow-roll parameters as
\begin{eqnarray}
\label{epss1}
{\varepsilon _{s1}} &\equiv& \frac{{{{\dot c}_s}}}{{H{c_s}}},
\\
\label{epssip1}
{\varepsilon _{s(i + 1)}} &\equiv& \frac{{{{\dot \varepsilon }_{si}}}}{{H{\varepsilon _{si}}}}.
\end{eqnarray}
Furthermore, using Eqs. (\ref{epsip1}), (\ref{dlnk}), (\ref{ns2}), (\ref{epss1}) and (\ref{epssip1}), we can obtain an expression for the running of the scalar spectral index as
\begin{equation}\label{dns}
\frac{{d{n_s}}}{{d\ln k}} =  - 2{\varepsilon _1}{\varepsilon _2} - {\varepsilon _2}{\varepsilon _3} - 3{\varepsilon _{s1}}{\varepsilon _{s2}}.
\end{equation}
If the sound speed is constant then the sound speed slow-roll parameters in Eqs. (\ref{ns2}) and (\ref{dns}) vanish and we recover the expressions for the scalar spectral index and the running of the scalar index that we expect in the standard inflationary scenario.

Now we turn to study the tensor perturbations in $f(T)$-gravity. According to \cite{Che11}, the equation governing the tensor perturbation $h_{ij}$ can be obtained as
\begin{equation}\label{d2hij}
{\ddot h_{ij}} + 3H{\dot h_{ij}} - \frac{{{\nabla^2}}}{{{a^2}}}{h_{ij}} + \gamma {\dot h_{ij}} = 0,
\end{equation}
where we have defined the parameter $\gamma$ as
\begin{equation}\label{gamma}
\gamma  \equiv \frac{{\dot T{f_{,TT}}}}{{{f_{,T}}}}.
\end{equation}
The tensor perturbation $h_{ij}$ is symmetric, transverse and traceless, i.e.
\begin{equation}\label{hij}
{h_{ij}} = {h_{ji}},\,\,\,\,\,\partial^ih_{ij} = 0,\,\,\,\,\,h_{ii} = 0.
\end{equation}
Due to these constraints, the tensor perturbation $h_{ij}$ has only two degrees of freedom which correspond to two polarizations of gravitational waves. We label the polarization state by $r$ and for each state we can write ${h_{ij}}(t,x)$ as a scalar field $h^r(t, x)$ multiplied by a polarization tensor $\xi _{ij}^r$ which is constant in space and time. Thus, the Fourier transformations of the tensor perturbation is given by
\begin{equation}\label{hijtx}
{h_{ij}}(t,x) = \sum\limits_{r = 1}^2 {\int {\frac{{{d^3}k}}{{{{(2\pi )}^{3/2}}}}} } {h^r}(t,k)\,\xi _{ij}^r\,{{\mathop{\rm e}\nolimits} ^{ikx}}.
\end{equation}
Using the above relation in Eq. (\ref{d2hij}), we get
\begin{equation}\label{d2hr}
{\ddot h^r} + \left( {3H + \gamma } \right){\dot h^r} + \frac{{{k^2}}}{{{a^2}}}{h^r} = 0.
\end{equation}
Applying the relation $d\tau  = dt/a$ in the above equation yields
\begin{equation}\label{d2hr2}
{h^r}'' + \frac{{2z_t'}}{{{z_t}}}{h^r}' + \frac{{{k^2}}}{{{a^2}}}{h^r} = 0,
\end{equation}
where the parameter $z_t$ is defined as it satisfies the differential equation
\begin{equation}\label{dzt}
\frac{{{{\dot z}_t}}}{{{z_t}}} = H + \frac{{\gamma}}{2}.
\end{equation}
The solution of this differential equation is
\begin{equation}\label{zt}
{z_t} = a\exp \left( { \int {\frac{\gamma }{2}\,dt} } \right).
\end{equation}
If we define the canonically normalized field
\begin{equation}\label{vrk}
v_k^r = \frac{{{z_t}}}{2}h^r{M_P},
\end{equation}
then Eq. (\ref{d2hr2}) gives rise to
\begin{equation}\label{d2vrk}
{v_k^r}'' + \left( {{k^2} - \frac{{z_t ''}}{{{z_t}}}} \right)v_k^r = 0.
\end{equation}
Following the same procedure used for the scalar perturbations, we can find the asymptotic solutions of Eq. (\ref{d2vrk}) and then we match the solutions at the horizon exit specified by $k=a H$. In this way, we get the tensor power spectrum in the framework of $f(T)$-gravity as
\begin{equation}\label{Pt0}
{{\cal P}_t} = 2{{\cal P}_h} = {\left. {\frac{{2{a^2}{H^2}}}{{{\pi ^2}M_P^2z_t^2}}} \right|_{k = aH}},
\end{equation}
which is sum of the power spectra ${{\cal P}_h}$ for two polarization modes of $h_{ij}$.

Here, we introduce the tensor-to-scalar ratio defined as
\begin{equation}\label{r}
r \equiv \frac{{{{\cal P}_t}}}{{{{\cal P}_s}}}.
\end{equation}
The tensor-to-scalar ratio is an important inflationary observable that is strictly constrained by the Planck 2015 observational data \cite{Pla15}. Hence, it can be used to distinguish viable inflationary models in light of the observational results. Another inflationary observable is the tensor spectral index defined as
\begin{equation}\label{nt}
{n_t} \equiv \frac{{d\ln {{\cal P}_t}}}{{d\ln k}},
\end{equation}
that specifies the scale dependence of the tensor power spectrum. The accuracy of current experimental devices is not adequate to measure the tensor spectral index but we may be able to determine it in the future.

In order to obtain a simpler relation for the tensor power spectrum in our $f(T)$-gravity model, it is convenient to define the parameter
\begin{equation}\label{delta}
\delta  \equiv \frac{{\left| \gamma  \right|}}{{2H}}.
\end{equation}
If $\delta  \ll 1$ then we can neglect the term $\gamma /2$ relative to the Hubble parameter $H$ in Eq. (\ref{dzt}) and therefore the resulting equation yields the standard solution $z_t=a$. Consequently, Eq. (\ref{Pt0}) reduces to the standard expression for the tensor power spectrum in the framework of Einstein gravity as
\begin{equation}\label{Pt}
{{\cal P}_t} = {\left. {\frac{{2{H^2}}}{{{\pi ^2}M_P^2}}} \right|_{k = aH}}.
\end{equation}
This relation must be calculated at the time of horizon crossing for which $k=a H$. This time is not exactly the same as the time of sound horizon crossing for which $c_s k=a H$, but to lowest order in the slow-roll parameters this difference is negligible \cite{Gar99}. Therefore, using Eqs. (\ref{Ps}), (\ref{r}) and (\ref{Pt}), the tensor-to-scalar ratio is obtained as
\begin{equation}\label{r2}
r = 16c_s^3{\varepsilon _1}.
\end{equation}
Also, using Eqs. (\ref{eps1}), (\ref{dlnk}), (\ref{nt}) and (\ref{Pt}), we can obtain the tensor spectral index as
\begin{equation}\label{nt2}
{n_t}=-2\varepsilon_1.
\end{equation}
From Eqs. (\ref{r2}) and (\ref{nt2}), we conclude that the inflationary observables are not independent and there is a so-called ``consistency relation'' between them as
\begin{equation}\label{rnt}
r =  - 8c_s^3{n_t}.
\end{equation}
For the case of $c_s=1$, this equation turns into the conventional consistency relation $r =  - 8{n_t}$ being valid in the standard inflationary framework based on the Einstein gravity. We see that the consistency relation in $f(T)$-gravity is different from the usual one in the standard inflationary model and therefore, in principle, inflation in $f(T)$-gravity is phenomenologically distinguishable from the standard inflationary model based on the Einstein gravity.

\section{Power-law inflation in $f(T)$-gravity}\label{secpl}

In the previous section, we studied the cosmological perturbations in the framework of $f(T)$-gravity and obtained the expressions for the inflationary observables. In this section, we will apply the obtained results for the power-law inflation. It has been shown that in the standard inflationary setting, the power-law inflation is driven by an exponential potential which is not favored in light of the recent observational data \cite{Rez15a, Unn13}. This motivates us to check viability of the power-law inflation in comparison with the Planck 2015 data in the framework of $f(T)$-gravity.

We consider the $f(T)$ function in the action (\ref{action}) to have the power-law form \cite{Wu10a, Lin10}
\begin{equation}\label{fT}
f\left( T \right) = T_0\left(\frac{T}{T_0}\right)^n,
\end{equation}
where $T_0$ and $n$ are constant. In the case of $n=1$, Eq. (\ref{fT}) recovers the TEGR, i.e. $f(T)= T$. From Eq. (\ref{cs}), we see that the $f(T)$ model (\ref{fT}) leads to a constant sound speed as
\begin{equation}\label{cs2}
c_s^2 = \frac{1}{2n - 1}.
\end{equation}
Here, due to having a physical speed of scalar perturbations, the sound speed should be real and subluminal, i.e. $0<c_s^2\leq 1$ \cite{Fra10}. This limits the parameter $n$ to be in the range of $n\geq 1$. In addition, from Eq. (\ref{cs2}) the sound speed slow-roll parameters (\ref{epss1}) and (\ref{epssip1}) vanish in our $f(T)$-gravity model (\ref{fT}).

In this section, we focus on the power-law inflation with the scale factor
\begin{equation}\label{apl}
a(t) = {a_i}{\left( {\frac{t}{{{t_i}}}} \right)^q},
\end{equation}
where $q>1$ is a constant parameter and $a_i$ is the scale factor of the universe at the initial time of inflation $t_i$. Throughout this paper, we normalize the scale factor of the universe relative to its value at the present time so that $a_0=1$. It should be noted that there is not a certain value for $a_i$ and $t_i$ because from Linde's idea of eternal inflation \cite{Lin86a, Lin86b}, we infer that the initial conditions of inflation are uncertain.

The power-law scale factor (\ref{apl}) yields the Hubble parameter as
\begin{equation}\label{Hpl}
H = \frac{q}{t}.
\end{equation}
Also, the first Hubble slow-roll parameter (\ref{eps1}) becomes
\begin{equation}\label{eps1pl}
{\varepsilon _1} = \frac{1}{q},
\end{equation}
which is constant and hence the other Hubble slow-roll parameters vanish. Since the first slow-roll parameter is constant and cannot reach unity, then inflation never ends and it is needed to introduce an additional reheating process to the final stages of inflation to make exit from the inflationary phase possible for this model. However, in \cite{Unn13} the authors has shown that invoking a non-canonical scalar fields can resolve this central drawback of the power-law inflation.

Within the framework of $f(T)$ model (\ref{fT}), we are interested in determining the form of the inflationary potential for the power-law inflation (\ref{apl}). For the case of $n=1$ which corresponds to the TEGR, i.e. $f(T)=T$, Eqs. (\ref{V}) and (\ref{phidot}) yield the potential and scalar field with respect to time as
\begin{eqnarray}
\label{Vtpln1}
V(t) &=& \frac{{3{q^2}}}{{{{\left( {{M_P}t} \right)}^2}}}M_P^4,
\\
\label{phitpln1}
\phi (t) &=& \sqrt {2q} \ln \left( {{M_P}t} \right){M_P},
\end{eqnarray}
respectively. Eliminating $t$ in the above equations, we find the inflationary potential as
\begin{equation}\label{Vphipln1}
V(\phi ) = 3q^2 e^{-\sqrt{\frac{2}{q}} \left( \frac{\phi}{M_P} \right)}M_P^4,
\end{equation}
which is the familiar result that we expect in the standard inflationary framework based on the Einstein gravity \cite{Rez15a, Unn13}.

Moreover, for $n>1$, from Eqs. (\ref{V}) and (\ref{phidot}) we obtain
\begin{eqnarray}
\label{Vtpl}
V( t ) &=& \frac{{\left( {2n - 1} \right)}}{2}\left( {\frac{{ - {T_0}}}{{M_P^2}}} \right){\left( {\frac{{6{q^2}}}{{ - {T_0}{t^2}}}} \right)^n}M_P^4,
\\
\label{phitpl}
\phi (t) &=& \frac{{{6^{n/2}}}}{{n - 1}}\sqrt {\frac{{n\left( {2n - 1} \right)q}}{3}} {\left( {\frac{q}{{\sqrt { - {T_0}} \,t}}} \right)^{n - 1}}{M_P},
\end{eqnarray}
respectively. Combining the two above equations, we obtain the inflationary potential driving the power-law inflation in our $f(T)$-gravity framework as
\begin{equation}\label{Vphipl}
V(\phi ) = {V_0}{\left( {\frac{\phi }{{{M_P}}}} \right)^m},
\end{equation}
where the parameters of $m$ and $V_0$ are given by
\begin{eqnarray}
\label{mpl}
m &=& \frac{{2n}}{{n - 1}},
\\
\label{V0pl}
{V_0} &=& \frac{2}{{\left( {2n - 1} \right)}}{\left( {\frac{{{{\left( {n - 1} \right)}^2}}}{{4nq}}} \right)^n}\left( { \frac{{{-T_0}}}{{M_P^2}}} \right)M_P^4.
\end{eqnarray}
As we see, in our $f(T)$-gravity scenario with $n>1$, the potential corresponding to the power-law inflation has a power-law form that this class of potentials includes the simplest chaotic inflationary models \cite{Lin83}, in which inflation starts from large values for the inflaton, $\phi  > {M_P}$.

The scalar power spectrum in our $f(T)$ model (\ref{fT}) with $n\geq 1$ is obtained from Eq. (\ref{Ps}) as
\begin{equation}\label{Pstpl}
{{\cal P}_s}(t) = {\left. {\frac{{{q^3}{{\left( {2n - 1} \right)}^{3/2}}}}{{8{\pi ^2}{{\left( {{M_P}t} \right)}^2}}}} \right|_{{c_s}k = aH}}.
\end{equation}
As we mentioned before, the power spectrum of the scalar perturbations should be evaluated at the sound horizon exit. Applying the relation $c_s k = aH$, we specify the time of the sound horizon exit as
\begin{equation}\label{tspl}
{t_s} = {\left( {\frac{{t_i^qk}}{{{a_i}q\sqrt {2n - 1} }}} \right)^{\frac{1}{{q - 1}}}}.
\end{equation}
Replacing this into Eq. (\ref{Pstpl}) gives the scalar power spectrum in terms of the comoving wavenumber $k$ as
\begin{equation}\label{Pskpl}
{{\cal P}_s}(k) = \frac{{{{\left( {2n - 1} \right)}^{3/2}}{q^3}}}{{8{\pi ^2}M_P^2}}{\left( {\frac{{{a_i}q\sqrt {2n - 1} }}{{t_i^qk}}} \right)^{\frac{2}{{q - 1}}}}.
\end{equation}
We see that in our $f(T)$ scenario, like the standard inflationary scenario, the power-law inflation leads to a power-law power spectrum for the scalar perturbations. Therefore, we can easily calculate the scalar spectral index from Eq. (\ref{ns}) as
\begin{equation}\label{nspl}
{n_s} = 1 - \frac{2}{{q - 1}},
\end{equation}
which does not have scale dependence so that it gives rise to a vanishing running of the scalar spectral index, $d{n_s}/d\ln k = 0$, which is in agreement with the Planck 2015 observational data at 68\% CL \cite{Pla15}. Also, the above result does not depend on the parameter $n$ and hence, the scalar spectral index in our $f(T)$ model is the same as one in the standard inflationary scenario.

Here, we are interested in investigating the tensor perturbations for the power-law inflation in our $f(T)$ model. First, we calculate the parameter $\delta$ from Eq. (\ref{delta}) and obtain
\begin{equation}\label{deltapl}
\delta  = \frac{{n - 1}}{q}.
\end{equation}
We see that if $q \gg n-1$ then $\delta \ll 1$. We assume this condition to be valid and we will verify it later in the present section. This assumption allows us to use Eq. (\ref{Pt}) for the tensor power spectrum that it leads to
\begin{equation}\label{Pttpl}
{{\cal P}_t}(t) = {\left. {\frac{{2{q^2}}}{{{\pi ^2}{{\left( {{M_P}t} \right)}^2}}}} \right|_{k = aH}}.
\end{equation}
We must calculate this at the time of horizon exit specified by the relation $k = aH$ as
\begin{equation}\label{ttpl}
{t_t} = {\left( {\frac{{t_i^qk}}{{{a_i}q}}} \right)^{\frac{1}{{q-1}}}}.
\end{equation}
Substituting Eq. (\ref{ttpl}) into Eq. (\ref{Pttpl}) gives
\begin{equation}\label{Ptkpl}
{{\cal P}_t}(k) = \frac{{2{q^2}}}{{{\pi ^2}M_P^2}}{\left( {\frac{{{a_i}q}}{{t_i^qk}}} \right)^{\frac{2}{{q - 1}}}}.
\end{equation}
With the help of above relation, Eq. (\ref{nt}) gives a scale-invariant tensor spectral index as
\begin{equation}\label{ntpl}
{n_t} =  - \frac{2}{{q - 1}}.
\end{equation}
We note that the tensor spectral index, like the scalar spectral index, does not depend on the parameter $n$ and therefore, it is identical to the one for the standard inflationary scenario. Also, using Eqs. (\ref{Pskpl}) and (\ref{Ptkpl}) in definition (\ref{r}), we find the tensor-to-scalar ratio as
\begin{equation}\label{rpl}
r = \frac{16}{q(2n - 1)^\frac{3q}{2(q - 1)}}.
\end{equation}

\begin{figure}[t]
\begin{center}
\scalebox{1}[1]{\includegraphics{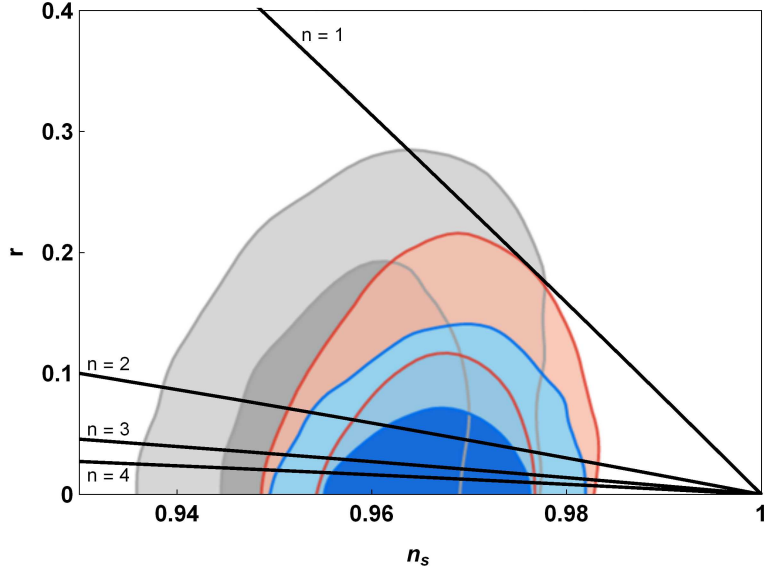}}
\caption{Prediction of the power-law inflation in $f(T)$-gravity in $r-n_s$ plane in comparison with the Planck 2015 results. The black lines indicate the predictions of the power-law inflation (\ref{apl}) in our $f(T)$ model (\ref{fT}) with different values of $n$. The marginalized joint 68\% and 95\% CL regions of Planck 2013, Planck 2015 TT+lowP and Planck 2015 TT,TE,EE+lowP data \cite{Pla15} are specified by grey, red and blue, respectively. The case of $n=1$ denotes the model $f (T ) = T$ corresponding
to the TEGR.}
\label{figrnspl}
\end{center}
\end{figure}

Now, we can use Eqs. (\ref{nspl}) and (\ref{rpl}) to plot the prediction of the power-law inflation in the $f(T)$ model (\ref{fT}) in $r-n_s$ plane. The plot is demonstrated in Fig. \ref{figrnspl} for different values of the parameter $n$ while the parameter $q$ is varying. Also, the marginalized joint 68\% and 95\% CL regions allowed by the Planck 2015 data \cite{Pla15} have been specified in the figure. We see that in contrary to the case of the TEGR ($n=1$), in our $f(T)$ model (\ref{fT}) with $n\gtrsim 2$, the prediction of the power-law inflation (\ref{apl}) can lie inside the 68\% CL region of Planck 2015 TT,TE,EE+lowP data \cite{Pla15}. Therefore, we conclude that in $f(T)$-gravity, the plower-law inflation can be resurrected in light of the Planck 2015 observational results.

Here, we check validity of the assumption $\delta \ll 1$ that we made before to use Eq. (\ref{Pt}) for the tensor power spectrum. From Eq. (\ref{nspl}), we see that a favored value for the scalar spectral index according to Planck 2015 TT,TE,EE+lowP data at 68\% CL ($n_s= 0.9644 \pm 0.0049$) \cite{Pla15}, is obtained for $q \approx 57$. Now, if we take $q \approx 57$ then from Eq. (\ref{deltapl}) for $n=2$, $3$ and $4$, we obtain $\delta \approx 0.02$, $0.04$ and $0.05$ which satisfy the condition $\delta \ll 1$ and consequently our assumption is valid.

In Eq. (\ref{Vphipl}) we found that the inflationary potential responsible for the power-law inflation in our $f(T)$ scenario has a power-law form. From Eq. (\ref{mpl}) we see that for $n \gtrsim 2$ the power $m$ for the inflationary potential varies in the range $2 < m \lesssim 4$. In the standard inflationary, the power-law potential (\ref{Vphipl}) with a power in this range, is not favored by Planck 2015 TT,TE,EE+lowP data \cite{Pla15}, as it has been shown in \cite{Rez15a}. But we see that in our $f(T)$ inflationary model, the power-law potential (\ref{Vphipl}) with $2 < m \lesssim 4$ can be consistent with Planck 2015 TT,TE,EE+lowP data \cite{Pla15} at 68\% CL. Interestingly enough, is that in our model, we can refine the self-interacting quartic potential $V(\phi ) \propto {\phi ^4}$ which has special reheating properties \cite{For87, Lid03}. This potential in our model corresponds to $n=2$ that can be compatible with the Planck 2015 results at 68\% CL while it is ruled out in the standard inflationary scenario \cite{Rez15a}. In \cite{Unn12, Bar14}, the authors have investigated some ideas to improve the prediction of the quartic potential in comparison with the observational data.

In the following, we estimate the inflationary observables in our model explicitly. We choose $n=2$ and $q=57$. As a result, from Eq. (\ref{nspl}), we find the scalar spectral index as $n_s=0.9643$ that is in agreement with Planck 2015 TT,TE,EE+lowP data at 68\% CL ($n_s= 0.9644 \pm 0.0049$) \cite{Pla15}. Also, using Eq. (\ref{rpl}), we see that our model predicts the tensor-to-scalar ratio as $r=0.0530$ which lies inside the 68\% CL region of Planck 2015 TT,TE,EE+lowP data \cite{Pla15} (see Fig. \ref{figrnspl}). From Eq. (\ref{ntpl}), the tensor spectral index is obtained as $n_t=-0.0357$ which satisfies the consistency relation (\ref{rnt}). The current experimental devices are not sufficiently accurate to measure the tensor spectral index $n_t$ with a suitable accuracy and the predicted value for this observable can be checked by more precise measurements in the future.

In the remaining of this section, we turn to check the validity of our discussion in the context of the $e$-folds number from the end of inflation. The $e$-folds number is used to express the amount of inflation and is defined as
\begin{equation}
\label{N}
N \equiv \ln \left({\frac{{{a_e}}}{a}} \right),
\end{equation}
where $a_e$ is the scale factor at the end of inflation. This definition is equivalent to
\begin{equation}
\label{dN}
dN =  - H dt.
\end{equation}
The CMB anisotropies correspond to the perturbations whose wavelengths crossed the Hubble
radius around $N_* \approx 50 - 60$ before the end of inflation \cite{Lid03, Dod03}. This result can be obtained with the assumption that during inflationary era, a slow-roll inflation has occurred that it leads to a quasi-de Sitter expansion of the universe with $H \approx {\rm{constant}}$. Furthermore, evolution of the universe after inflation is assumed to be governed by the standard model of cosmology. In the present work, we have used these two assumptions, hence we can take the $e$-folds number corresponding to the horizon crossing as $N_* \approx 50 - 60$ from the end of inflation. Since inflation with the power-law scale factor (\ref{apl}) cannot end by slow-roll violation, we follow the logic of \cite{Mar14} and introduce an extra parameter $t_e$ related to the time in which an unspecified reheating mechanism is triggered to end of inflation. Now, we can solve the differential equation (\ref{dN}) for the power-law scale factor (\ref{apl}) and get
\begin{equation}\label{tNpl}
t = {t_e}{e^{ - N/q}},
\end{equation}
where we have used the initial condition ${N_e} \equiv N({t_e}) = 0$ from Eq. (\ref{N}).

Substituting Eq. (\ref{tNpl}) into (\ref{Pstpl}), one can get the scalar power spectrum in terms of the $e$-folds number as
\begin{equation}\label{PsNpl}
{{\cal P}_s} (N)= \frac{{{{\left( {2n - 1} \right)}^{3/2}}{q^3}}}{{8{\pi ^2}{{\left( {{M_P}{t_e}} \right)}^2}}}{e^{2N/q}}.
\end{equation}
To find an expression for the scalar spectral index $n_s$, we note that in the relation $a H=c_s k$, the Hubble parameter $H$ is approximately constant during slow-roll inflation, and also the sound speed $c_s$ is constant for our $f(T)$-gravity model. Consequently, by use of Eq. (\ref{dN}), we will have
\begin{equation}\label{dlnkdN}
d\ln k \approx  - dN,
\end{equation}
which is valid around the sound horizon exit. Therefore, using Eqs. (\ref{PsNpl}) and (\ref{dlnkdN}) in (\ref{ns}), we obtain
\begin{equation}\label{nsNpl}
{n_s} = 1 - \frac{2}{q},
\end{equation}
which is independent of the $e$-folds number and consequently gives a vanishing running of the scalar spectral index, $d{n_s}/d\ln k = 0$. This is in agreement with our previous result obtained by use of the comoving wavenumber.

Now, we want to get the tensor power spectrum with respect to the $e$-folds number. We noticed before that the tensor power spectrum should be evaluated at the time of horizon exit where $a H= k$. To lowest order in the slow-roll parameters, we can ignore the difference between the time of horizon exit and the sound horizon exit in the slow-roll regime \cite{Gar99}. Therefore, we can insert Eq. (\ref{tNpl}) into (\ref{Pttpl}) and reach
\begin{equation}\label{PtNpl}
{{\cal P}_t}(N) = \frac{{2{q^2}}}{{{\pi ^2}{{\left( {{M_P}{t_e}} \right)}^2}}}{e^{2N/q}}.
\end{equation}
Here, we can use Eqs. (\ref{PsNpl}) and (\ref{PtNpl}) in (\ref{r}) and obtain the tensor-to-scalar ratio as
\begin{equation}\label{rNpl}
r = \frac{{16}}{{{{\left( {2n - 1} \right)}^{3/2}}q}}.
\end{equation}
Moreover, with the help of Eqs. (\ref{nt}), (\ref{dlnkdN}) and (\ref{PtNpl}), the tensor spectral index reads
\begin{equation}\label{ntNpl}
{n_t} =  - \frac{2}{q}.
\end{equation}
We see that the inflationary observables $n_s$, $r$ and $n_t$ for the power-law inflation become independent of the $e$-folds number in our $f(T)$-gravity scenario. This situation is similar to that of the standard inflationary model where these quantities depend only on the parameter $q$ appeared in the power-law scale factor $a\propto t^q$ \cite{Mar14, Unn13}. Now, we estimate these inflationary observables again and compare them with our previous results obtained by use of the comoving wavenumber. For this purpose, we take $n=2$ and $q=57$ again, and find ${n_s} = 0.9649$, $r = 0.0540$ and ${n_t} =  - 0.0351$ from Eqs. (\ref{nsNpl}), (\ref{rNpl}) and (\ref{ntNpl}), respectively. We see that these results are very close to those found before by use of the comoving wavenumber. This verifies the validity of our discussion in the context of $e$-folds number for the power-law inflation in our $f(T)$-gravity model. The small deviations in the results obtained by use of the $e$-folds number relative to those calculated by use of the comoving wavenumber arise from two reasons. The first reason is that in Eq. (\ref{dlnkdN}), we neglected from the slow-roll Hubble parameters relative to unity while their contributions are included in the calculations on the base of the comoving wavenumber. The second one is that to get Eq. (\ref{PtNpl}), we ignored the difference between the times of horizon exit for the scalar and tensor perturbations but this difference is considered in the calculations based on the comoving wavenumber.

\section{Intermediate inflation in $f(T)$-gravity}\label{secint}

In this section, we are interested to study the intermediate inflation in the framework of $f(T)$-gravity. The scale factor of intermediate inflation takes the form
\begin{equation}\label{aint}
a(t) = a_i \exp \left[ {A{{\left( {{M_P}t} \right)}^\lambda}}
\right],
\end{equation}
where $A>0$ and $0<\lambda<1$ \cite{Bar06, Bar07, Rez15b}. Furthermore, $a_i$ indicates the scale factor at the initial time of inflation. This scale factor leads to the Hubble parameter as
\begin{equation}\label{Hint}
H = \frac{{A\lambda }}{{{{\left( {{M_P}t} \right)}^{1 - \lambda }}}}{M_P}.
\end{equation}
In addition, the first Hubble slow-roll parameter (\ref{eps1}) reads
\begin{equation}\label{eps1int}
{\varepsilon _1} = \frac{{\left( {1 - \lambda } \right)}}{{A\lambda {{\left( {M_Pt} \right)}^\lambda }}},
\end{equation}
which is a decreasing function during inflation. Therefore, like the power-law inflation, the intermediate inflation cannot end without introducing an additional reheating process to the final stages of the inflationary phase. An idea to overcome the end of intermediate inflation problem has been proposed in \cite{Rez15b}.

Here, we want to determine the inflationary potential corresponding to the intermediate inflation in our model. We first concentrate on the case of $n=1$ for which our $f(T)$ model (\ref{fT}) recovers the TEGR, i.e. $f(T)=T$. For this case, Eqs. (\ref{V}) and (\ref{phidot}) lead to
\begin{eqnarray}
\label{Vtintn1}
V(t) &=& \frac{{3{A^2}{\lambda ^2}}}{{{{\left( {{M_P}t} \right)}^{2\left( {1 - \lambda } \right)}}}}M_P^4,
\\
\label{phitintn1}
\phi (t) &=& 2\sqrt {\frac{{2A\left( {1 - \lambda } \right)}}{\lambda }} {\left( {{M_P}t} \right)^{\lambda /2}}{M_P},
\end{eqnarray}
that can be combined to result in the inverse power-law inflationary potential
\begin{equation}\label{Vphiintn1}
V(\phi ) = 3{\left( {A\lambda } \right)^2}{\left[ {\frac{{8A\left( {1 - \lambda } \right)}}{\lambda }} \right]^{\frac{{2\left( {1 - \lambda } \right)}}{\lambda }}}{\left( {\frac{\phi }{{{M_P}}}} \right)^{ - \frac{{4\left( {1 - \lambda } \right)}}{\lambda }}}M_P^4,
\end{equation}
which we expect for the intermediate inflation in the framework of Einstein gravity \cite{Bar06, Bar07, Rez15b}.

For $n>1$, using Eqs. (\ref{V}) and (\ref{phidot}), we obtain the inflationary potential and scalar field in terms of the cosmic time $t$ as
\begin{eqnarray}
\label{Vtint}
V(t) &=& \frac{{\left( {2n - 1} \right)}}{2}{\left( { \frac{{M_P^2}}{{{-T_0}}}} \right)^{n - 1}}{\left[ {\frac{{\sqrt 6 A\lambda }}{{{{\left( {{M_P}t} \right)}^{1 - \lambda }}}}} \right]^{2n}}M_P^4,
\\
\label{phitint}
\phi (t) &=& \frac{{2{{\left[ {2n\left( {2n - 1} \right)A\lambda \left( {1 - \lambda } \right)} \right]}^{1/2}}{{\left[ {\sqrt 6 A\lambda \left( {\frac{{{M_P}}}{{\sqrt { - {T_0}} }}} \right)} \right]}^{n - 1}}}}{{\left[ {2n\left( {1 - \lambda } \right) + \lambda  - 2} \right]{{\left( {{M_P}t} \right)}^{n(1 - \lambda ) + \frac{\lambda }{2} - 1}}}}{M_P},
\end{eqnarray}
respectively. Eliminating $t$ between these equations gives the inflationary potential as
\begin{equation}\label{Vphiint}
V(\phi ) = {V_0}{\left( {\frac{\phi }{{{M_P}}}} \right)^m},
\end{equation}
where the parameters of $m$ and $V_0$ are defined as
\begin{eqnarray}
\label{mint}
m &=& \frac{{4n\left( {1 - \lambda } \right)}}{{2n\left( {1 - \lambda } \right) + \lambda  - 2}},
\\
\label{V0int}
{V_0} &=& M_P^4{\left[ {\frac{{{2^{\frac{{n\left( {8 - 7\lambda } \right) + \lambda  - 2}}{{2n}}}}{3^{\lambda /2}}{{\left( {2n - 1} \right)}^{\frac{{2 - \lambda }}{{2n}}}}A\lambda }}{{{{\left( {\frac{{{{\left[ {2n(\lambda  - 1) - \lambda  + 2} \right]}^2}}}{{n\left( {1 - \lambda } \right)}}} \right)}^{1 - \lambda }}{{\left( {\frac{{ - {T_0}}}{{M_P^2}}} \right)}^{\frac{{\left( {n - 1} \right)\left( {2 - \lambda } \right)}}{{2n}}}}}}} \right]^{\frac{{2n}}{{2n(\lambda  - 1) - \lambda  + 2}}}}.
\end{eqnarray}
We see that in our $f(T)$ model (\ref{fT}) with $n>1$, the potential responsible for the intermediate inflation, like the one driving the power-law inflation, has the power-law form (\ref{Vphiint}), but with a different expression for the power $m$.

Now, we are in position to obtain the scalar power spectrum in our $f(T)$ scenario (\ref{fT}) with $n\geq1$. From Eq. (\ref{Ps}) we have
\begin{equation}\label{Pstint}
{{\cal P}_s}(t) = {\left. {\frac{{{{\left( {2n - 1} \right)}^{3/2}}{{\left( {A\lambda } \right)}^3}}}{{8{\pi ^2}\left( {1 - \lambda } \right){{\left( {{M_P}t} \right)}^{2 - 3\lambda }}}}} \right|_{{c_s}k = aH}}.
\end{equation}
From the relation $c_s k = aH$, the time of sound horizon exit reads
\begin{equation}\label{tsint}
{t_s} = {\left\{ {\frac{{\lambda  - 1}}{{A\lambda }}{W_{ - 1}}\left[ {\frac{{A\lambda }}{{\lambda  - 1}}{{\left( {\frac{k}{{{a_i}A\lambda \sqrt {2n - 1} {M_P}}}} \right)}^{\frac{\lambda }{{\lambda  - 1}}}}} \right]} \right\}^{1/\lambda }}M_P^{ - 1},
\end{equation}
where $W$ is the Lambert function defined as solution of the equation $y{{\mathop{\rm e}\nolimits} ^y} = x$ \cite{Cor96}. In the complex plane, the equation $y{{\mathop{\rm e}\nolimits} ^y} = x$ has a countably infinite number of solutions which are denoted by $W_k(x)$ with $k$ varying over the integers. For all real $x \ge 0$, the equation has exactly one real solution which is represented by $y=W(x) \equiv W_0(x)$. For all real $x$ in the range $x<0$, there are exactly two real solutions. The larger one is represented by $y=W(x)$ while the smaller one is denoted by $y=W_{-1}(x)$.

Substituting Eq. (\ref{tsint}) into (\ref{Pstint}), the scalar power spectrum is obtained in terms of the comoving wavenumber $k$ as
\begin{equation}\label{Pskint}
{{\cal P}_s}(k) = \frac{{{{\left( {A\lambda } \right)}^3}{{\left( {2n - 1} \right)}^{3/2}}}}{{8{\pi ^2}\left( {1 - \lambda } \right)}}{\left\{ {\frac{{\lambda  - 1}}{{A\lambda }}{W_{ - 1}}\left[ {\frac{{A\lambda }}{{\lambda  - 1}}{{\left( {\frac{k}{{{a_i}A\lambda \sqrt {2n - 1} {M_P}}}} \right)}^{\frac{\lambda }{{\lambda  - 1}}}}} \right]} \right\}^{\frac{{3\lambda  - 2}}{\lambda }}}.
\end{equation}
Now, we can use the above result in Eq. (\ref{ns}) and get the scalar spectral index as
\begin{equation}\label{nsint}
{n_s} = 1 + \frac{{2 - 3\lambda }}{{1 - \lambda }}{\left\{ {{W_{ - 1}}\left[ {\frac{{A\lambda }}{{\lambda  - 1}}{{\left( {\frac{k}{{{a_i}A\lambda \sqrt {2n - 1} {M_P}}}} \right)}^{\frac{\lambda }{{\lambda  - 1}}}}} \right] + 1} \right\}^{ - 1}}.
\end{equation}
The above equation yields the running of the scalar spectral index as
\begin{equation}\label{dnsint}
\frac{{d{n_s}}}{{d\ln k}} = \frac{{\lambda \left( {2 - 3\lambda } \right){W_{ - 1}}\left[ {\frac{{A\lambda }}{{\lambda  - 1}}{{\left( {\frac{k}{{{a_i}A\lambda \sqrt {2n - 1} {M_P}}}} \right)}^{\frac{\lambda }{{\lambda  - 1}}}}} \right]}}{{{{\left( {1 - \lambda } \right)}^2}{{\left\{ {{W_{ - 1}}\left[ {\frac{{A\lambda }}{{\lambda  - 1}}{{\left( {\frac{k}{{{a_i}A\lambda \sqrt {2n - 1} {M_P}}}} \right)}^{\frac{\lambda }{{\lambda  - 1}}}}} \right] + 1} \right\}}^3}}}.
\end{equation}

In order to study the tensor perturbations for the intermediate inflation in our $f(T)$-gravity scenario, first we note that Eq. (\ref{delta}) leads to
\begin{equation}\label{deltaint}
\delta  = \frac{{\left( {n - 1} \right)\left( {1 - \lambda } \right)}}{{A\lambda {{\left( {{M_P}t} \right)}^{ - \lambda }}}},
\end{equation}
which is suppressed as time last during inflation. Therefore, it makes sense to suppose that $\delta \ll 1$ at the time of horizon exit. We will verify the validity of this assumption later in this section. The assumption $\delta \ll 1$ allows us to use Eq. (\ref{Pt}) for the tensor power spectrum that it leads to
\begin{equation}\label{Pttint}
{{\cal P}_t}(t) = {\left. {\frac{{2{{\left( {A\lambda } \right)}^2}}}{{{\pi ^2}{{\left( {MPt} \right)}^{2\left( {1 - \lambda } \right)}}}}} \right|_{k = aH}}.
\end{equation}
The relation $k = aH$ results in the time of horizon exit to be
\begin{equation}\label{ttint}
{t_t} = {\left\{ {\frac{{\lambda  - 1}}{{A\lambda }}{W_{ - 1}}\left[ {\frac{{A\lambda }}{{\lambda  - 1}}{{\left( {\frac{k}{{{a_i}A\lambda {M_P}}}} \right)}^{\frac{\lambda }{{\lambda  - 1}}}}} \right]} \right\}^{1/\lambda }}M_P^{ - 1}.
\end{equation}
Using this in Eq. (\ref{Pttint}), we will have
\begin{equation}\label{Ptkint}
{{\cal P}_t}(k) = \frac{{2{A^2}{\lambda ^2}}}{{{\pi ^2}}}{\left\{ {\frac{{\lambda  - 1}}{{A\lambda }}{W_{ - 1}}\left[ {\frac{{A\lambda }}{{\lambda  - 1}}{{\left( {\frac{k}{{{a_i}A\lambda {M_P}}}} \right)}^{\frac{\lambda }{{ - 1 + \lambda }}}}} \right]} \right\}^{\frac{{2\left( {\lambda  - 1} \right)}}{\lambda }}}.
\end{equation}
From the above result together with Eq. (\ref{nt}), one finds
\begin{equation}\label{ntint}
{n_t} = 2{\left\{ {{W_{ - 1}}\left[ {\frac{{A\lambda }}{{\lambda  - 1}}{{\left( {\frac{k}{{{a_i}A\lambda {M_P}}}} \right)}^{\frac{\lambda }{{\lambda  - 1}}}}} \right] + 1} \right\}^{ - 1}}.
\end{equation}
Also, the tensor-to-scalar ratio results from Eqs. (\ref{Pskint}) and (\ref{Ptkint}) as
\begin{equation}\label{rint}
r = \frac{{16{{\left\{ { - {W_{ - 1}}\left[ {\frac{{A\lambda }}{{\lambda  - 1}}{{\left( {\frac{k}{{{a_i}A\lambda \sqrt {2n - 1} {M_P}}}} \right)}^{\frac{\lambda }{{\lambda  - 1}}}}} \right]} \right\}}^{\frac{{2 - 3\lambda }}{\lambda }}}}}{{{{\left( {2n - 1} \right)}^{3/2}}{{\left\{ { - {W_{ - 1}}\left[ {\frac{{A\lambda }}{{\lambda  - 1}}{{\left( {\frac{k}{{{a_i}A\lambda {M_P}}}} \right)}^{\frac{\lambda }{{\lambda  - 1}}}}} \right]} \right\}}^{\frac{{2\left( {1 - \lambda } \right)}}{\lambda }}}}}.
\end{equation}

\begin{figure}[t]
\begin{center}
\scalebox{1}[1]{\includegraphics{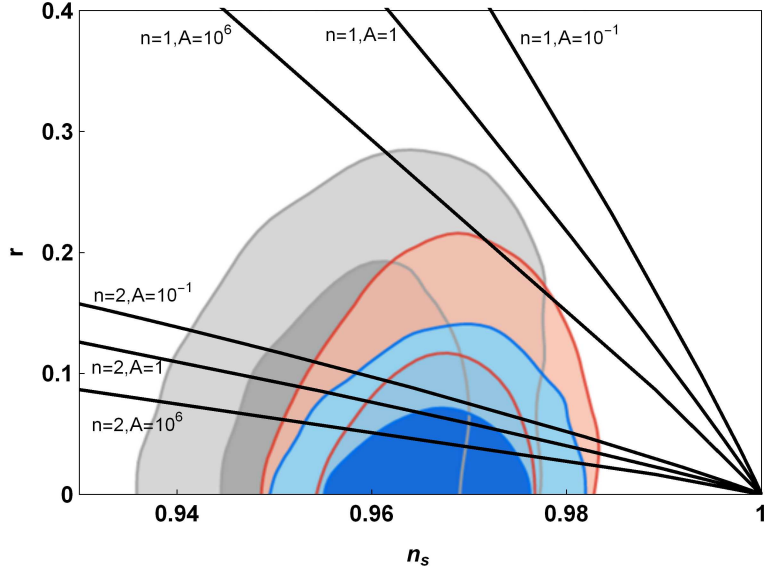}}
\caption{Same as Fig. \ref{figrnspl}, but for the intermediate inflation (\ref{aint}) in our $f(T)$-gravity scenario (\ref{fT}) with different values of $n$ and $A$. }
\label{figrnsint}
\end{center}
\end{figure}

So far, we have calculated the relations corresponding to the inflationary observables in terms of the comoving wavenumber. Now, we are able to check the viability of our inflationary model in light of the Planck 2015 results. We calculate the inflationary observables at the pivot scale ${k_*} = 0.05\,{\rm{Mpc}^{ - 1}}$ as adopted by the Planck 2015 collaboration \cite{Pla15}. We fix the scalar power spectrum in Eq. (\ref{Pskint}) at the pivot scale as $\ln \left[ {{{10}^{10}}{{\cal P}_s}\left( {{k_*}} \right)} \right] = 3.094$ from Planck 2015 TT,TE,EE+lowP data combination \cite{Pla15}. As a result, we will have an equation that results in a value for the parameter $a_i$ for each set of the parameters $n$, $A$ and $\lambda$. Consequently, we can plot the $r-n_s$ diagram for our model by use of Eqs. (\ref{nsint}) and (\ref{rint}) for different values of $n$ and $A$ while $\lambda$ is varying in the range $0 < \lambda  < 1$. This diagram is shown in Fig. \ref{figrnsint} and also the marginalized joint 68\% and 95\% CL regions allowed by the Planck 2015 data \cite{Pla15} have been specified in the figure. In Fig. \ref{figrnsint}, we see that the standard intermediate inflation based on the Einstein gravity ($n=1$) is disfavored in light of the Planck 2015 results. This result is in agreement with \cite{Rez15a, Rez15b}. However, if we choose $n \gtrsim 2$ then the intermediate inflation in our $f(T)$-gravity scenario can be consistent with the Planck 2015 data. For instance, for $n = 2$ and $A=1$ (or $A=10^6$), the prediction of our model can lie within the 68\% CL region favored by Planck 2015 TT,TE,EE+lowP data \cite{Pla15}.

\begin{figure}[t]
\begin{center}
\scalebox{1}[1]{\includegraphics{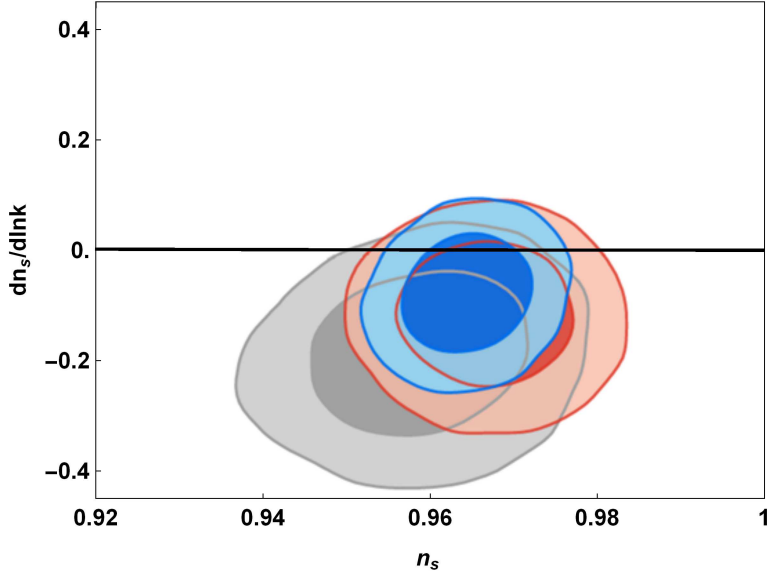}}
\caption{Prediction of the intermediate inflation (\ref{aint}) in our $f(T)$-gravity scenario (\ref{fT}) in $d{n_s}/d\ln k-{n_s}$ plane in comparison with the Planck 2015 results. The prediction of our model with $n=2$ and $A=1$ is shown by a black line. The grey, red and blue marginalized joint 68\% and 95\% CL regions correspond to Planck 2013, Planck 2015 TT+lowP and Planck 2015 TT,TE,EE+lowP data \cite{Pla15}, respectively.}
\label{fignsdns}
\end{center}
\end{figure}

Now, we test the prediction of our model in the $d{n_s}/d\ln k-{n_s}$ plane in comparison with the observational results of Planck 2015. To this aim, we consider $n=2$ and $A=1$. Subsequently, we use Eqs. (\ref{nsint}) and (\ref{dnsint}) to plot $d{n_s}/d\ln k$ versus $n_s$. The plot is represented in Fig. \ref{fignsdns} and we conclude that the prediction of intermediate inflation in our $f(T)$-gravity scenario can lie inside the joint 68\% CL region of Planck 2015 TT,TE,EE+lowP data \cite{Pla15}.

In the following, we proceed to estimate the inflationary observables in our model explicitly. We choose $n=2$ and $A=1$. With this selection, if the parameter $\lambda$ varies in the range $0.321 \lesssim \lambda \lesssim 0.329$, then the result of intermediate inflation can be placed within the joint 68\% CL region favored by Planck 2015TT,TE,EE+lowP data \cite{Pla15}. For the aforementioned range of $\lambda$, the parameter $\delta$ varies in the interval $0.020 \lesssim \delta \lesssim 0.026$, where we have used Eq. (\ref{deltaint}) for $\delta$ and it has been evaluated at the time of horizon exit given by Eq. (\ref{ttint}). Therefore, the assumption $\delta \ll 1$ which we have used before to apply Eq. (\ref{Pttint}) for the tensor power spectrum, is valid.

It should be noted that as the parameter $\lambda$ approaches $2/3$, the scalar power spectrum approaches the scale-invariant Harrison-Zel'’dovich spectrum with ${n_s} = 1$ that it is ruled out by the Planck 2015 results \cite{Pla15}. For the values of $\lambda$ in the interval $2/3<\lambda<1$, we will have a blue-tilted spectrum ($n_s>1$) that it is also ruled out by the Planck 2015 data \cite{Pla15}. However, we choose $\lambda=0.325$ in what follows. As a result, we find the inflationary observables from Eqs. (\ref{nsint}), (\ref{dnsint}) and (\ref{rint}) as ${n_s} = 0.9646$, $d{n_s}/d\ln k = 0.0004$ and $r = 0.0684$, respectively, that they are inside the joint 68\% CL region for Planck 2015TT,TE,EE+lowP data \cite{Pla15}. Additionally, Eq. (\ref{ntint}) gives the tensor spectral index as ${n_t} = -0.0463$ which also satisfies the consistency relation (\ref{rnt}).

In Eq. (\ref{Vphiint}) we found that within the framework of our $f(T)$ model (\ref{fT}), the inflationary potential corresponding to the intermediate inflation takes the power-law form $V(\phi ) = {V_0}{\left( {\phi /{M_P}} \right)^m}$ in which the power $m$ is determined by Eq. (\ref{mint}). For the allowed range $0 < \lambda  < 2/3$, from Eq. (\ref{mint}) the parameter $m$ varies in the interval $\frac{{2n}}{{n - 1}} < m < \frac{{2n}}{{n - 2}}$. In the case of our study, if we take $n=2$ and $\lambda=0.325$ then Eq. (\ref{mint}) gives the power $m=5.268$. We see that in our $f(T)$-gravity model (\ref{fT}), the prediction of the power-law potential (\ref{Vphiint}) with $m=5.268$ can be in agreement with the Planck 2015 data at 68\% CL, while in the standard inflationary scenario, this potential is disfavored by the observational results \cite{Rez15a}.

In what follows, we again verify validity of our results for the inflationary observables by use of the $e$-folds number from the end of inflation. We note that inflation with the intermediate scale factor (\ref{aint}) cannot stop by slow-roll violation. To overcome this problem, we again follow the approach of \cite{Mar14} and use the additional parameter $t_e$ which refers to the time in which an unknown reheating process begins to happen to stop inflation. In this way, we can solve the differential equation (\ref{dN}) for the intermediate scale factor (\ref{aint}) and find
\begin{equation}\label{tNint}
t = {\left[ {{{\left( {{M_P}{t_e}} \right)}^\lambda } - \frac{N}{A}} \right]^{1/\lambda }}M_P^{ - 1},
\end{equation}
where we have used the initial condition ${N_e} \equiv N({t_e}) = 0$ from Eq. (\ref{N}).

If we insert $t$ from Eq. (\ref{tNint}) into (\ref{Pstint}), the scalar power spectrum turns into
\begin{equation}\label{PsNint}
{{\cal P}_s}(N) = \frac{{{A^{2/\lambda }}{\lambda ^3}{{\left( {2n - 1} \right)}^{3/2}}}}{{8{\pi ^2}\left( {1 - \lambda } \right)}}{\left[ {A{{\left( {{M_P}{t_e}} \right)}^\lambda } - N} \right]^{\frac{{3\lambda  - 2}}{\lambda }}}.
\end{equation}
The above equation together with Eqs. (\ref{ns}) and (\ref{dlnkdN}) give rise to
\begin{equation}\label{nsNint}
{n_s} = \frac{{2 + \lambda \left[ {A{{\left( {{M_P}{t_e}} \right)}^\lambda } + 3 - N} \right]}}{{\lambda \left[ {A{{\left( {{M_P}{t_e}} \right)}^\lambda } - N} \right]}}.
\end{equation}
Also, we can apply Eq. (\ref{dlnkdN}) for the above relation which yields
\begin{equation}\label{dnsNint}
\frac{{d{n_s}}}{{d\ln k}} = \frac{{2 - 3\lambda }}{{\lambda {{\left[ {A{{\left( {{M_P}{t_e}} \right)}^\lambda } - N} \right]}^2}}}.
\end{equation}
In order to get the tensor power spectrum in terms of the $e$-folds number, we note that we can ignore the difference between the times of horizon exit for the scalar and tensor perturbations,
as discussed in the previous section. Consequently, we can use Eq. (\ref{tNint}) in (\ref{Pttint}) and obtain
\begin{equation}\label{PtNint}
{{\cal P}_t}(N) = \frac{{2{A^{2/\lambda }}{\lambda ^2}}}{{{\pi ^2}}}{\left[ {A{{\left( {{M_P}{t_e}} \right)}^\lambda } - N} \right]^{ - \frac{{2\left( {1 - \lambda } \right)}}{\lambda }}}.
\end{equation}
Substituting Eqs. (\ref{PsNint}) and (\ref{PtNint}) into (\ref{r}), one can get
\begin{equation}\label{rNint}
r = \frac{{16\left( {1 - \lambda } \right)}}{{{{\left( {2n - 1} \right)}^{\frac{3}{2}}}\lambda \left[ {A{{\left( {{M_P}{t_e}} \right)}^\lambda } - N} \right]}}.
\end{equation}
Furthermore, by use of Eqs. (\ref{dlnkdN}) and (\ref{PtNpl}) in (\ref{nt}), we find
\begin{equation}\label{ntNint}
{n_t} =  - \frac{{2\left( {1 - \lambda } \right)}}{{\lambda \left[ {A{{\left( {{M_P}{t_e}} \right)}^\lambda } - N} \right]}}.
\end{equation}
So far, we obtained the inflationary observables in terms of the $e$-folds number $N$. To compare these observables with observation, we should evaluate them at the horizon exit corresponding to the $e$-folds number $N_* \approx 50 - 60$. To determine the parameter $t_e$ in terms of the other parameters of the model, we fix the amplitude of the scalar power spectrum in Eq. (\ref{PsNint}) as $\ln \left[ {{{10}^{10}}{{\cal P}_s}\left( {{N_*}} \right)} \right] = 3.094$ from Planck 2015 TT,TE,EE+lowP data \cite{Pla15}. Now, we want to estimate the inflationary observables again and compare them with our previous results calculated by use of the comoving wavenumber. As before, we consider $n=2$, $A=1$ and $\lambda=0.325$. Now if we take $N_*=60$, then by use of Eqs. (\ref{nsNint}), (\ref{dnsNint}), (\ref{rNint}) and (\ref{ntNint}), we obtain ${n_s} = 0.9654$, $d{n_s}/d\ln k = 0.0004$, $r = 0.0702$ and ${n_t} =  - 0.0456$, respectively. Notice that for $N_*=50$, the results are very close to those obtained for $N_*=60$. We conclude that the results obtained by use of the $e$-folds number are approximately close to those obtained before by use of the comoving wavenumber. This confirms the validity of our study in the context of $e$-folds number for the intermediate inflation in our $f(T)$-gravity scenario.

\section{Conclusions}\label{seccon}

Here, we first represented a brief review on the background cosmology in $f(T)$-gravity in the presence of a canonical scalar field. Then, we studied the cosmological perturbations in the framework of $f(T)$-gravity. We obtained the necessary equations governing the scalar and tensor perturbations and solved them to find the scalar and tensor power spectra. Subsequently, we obtained the relations of the inflationary observables for our model. We found that the consistency relation for the inflationary model based on $f(T)$-gravity is different from the one for the standard inflationary model based on the Einstein gravity. Consequently, in principle, inflation in $f(T)$-gravity is phenomenologically distinguishable from the standard inflationary model based on Einstein's general relativity.

Next, we considered the $f(T)$ function in the action to have the power-law form $f(T) = {T_0}{(T/{T_0})^n}$ where $n \geq 1$. For $n=1$, the TEGR is recovered, i.e. $f(T)=T$. Then, we investigated the power-law inflation characterized by the scale factor $a(t) \propto {t^q}$ where $q>1$. In the Einstein gravity, the power-law inflation arises from the exponential potential $V(\phi ) \propto \exp \left[ { - \sqrt {2/q} \left( {\phi /{M_P}} \right)} \right]$ that is not favored in light of the Planck 2015 data. But, in our inflationary setting based on $f(T)$-gravity, if we choose $n \gtrsim 2$ then the power-law inflation can be in agreement with the Planck 2015 results at 68\% CL. In our scenario, the power-law inflation arises from the power-law potential $V(\phi ) \propto {\phi ^m}$ where $m=2n/(n-1)$. For $n \gtrsim 2$, the power $m$ varies in the range $2 < m \lesssim 4$. In the standard inflationary model, the power-law potential with this range of $m$ is not favored according to the Planck 2015 data while in our inflationary model, this potential can be consistent with the observational data. Interestingly enough, is that the self-interacting quartic potential $V(\phi ) \propto {\phi ^4}$ which has special reheating properties, can be consistent with the Planck 2015 data in our $f(T)$-gravity scenario while it is ruled out in the standard inflationary setting.

Within the inflationary framework of $f(T)$-gravity, we also examined the intermediate inflation with the scale factor $a(t) \propto \exp \left[ {A{{\left( {{M_P}t} \right)}^\lambda }} \right]$, where $A>0$ and $0 < \lambda <1$. In the standard inflationary scenario based on the Einstein gravity, the intermediate inflation is driven by the inverse power-law potential $V(\phi ) \propto {\phi ^{ - 4(1 - \lambda )/\lambda }}$ that it is not compatible with the Planck 2015 data. But in our $f(T)$-gravity model, the potential responsible for the intermediate inflation takes the power-law form $V(\phi ) \propto {\phi ^m}$ where $m{\rm{  = }}4n\left( {1 - \lambda } \right)/\left[ {2n\left( {1 - \lambda } \right) + \lambda  - 2} \right]$. We found that for $n \gtrsim 2$ and $0 < \lambda  < 2/3$, the intermediate inflation driven by the power-law potential in our $f(T)$-gravity scenario, can be consistent with the Planck 2015 results.

We further checked the validity of our discussion in the context of the $e$-folds number for the power-law and intermediate inflations in our $f(T)$-gravity model. Using the $e$-folds number, we computed the inflationary observables and found that their values are close to those obtained before by use of the comoving wavenumber.

\subsection*{Acknowledgements}

The authors thank the referee for his or her valuable comments. The work of A. Abdolmaleki has been supported financially by Research Institute for Astronomy and Astrophysics of
Maragha (RIAAM) under research project No. 1/4165-115.

\end{document}